\documentstyle[aps,floats,twocolumn,epsfig]{revtex}

\begin{document} 

\wideabs{
\title{An all-cryogenic THz transmission spectrometer}

\author{P.J. Burke,$^{a)}$ J.P. Eisenstein}

\address{Condensed Matter Physics, Caltech, Pasadena, CA  91125}

\author{L.N. Pfeiffer, K.W. West}

\address{Bell Laboratories, Lucent Technologies, Murray Hill, NJ 07974}
\date{\today} 
\maketitle

\begin{abstract}

This paper describes a THz transmission spectrometer for the spectral 
range of 2-65~cm$^{-1}$ (100~GHz~to~2~THz) with a spectral resolution 
of at least 1.8~cm$^{-1}$ (50~GHz)
where the source, sample, and detector are all fully contained in a cryogenic
environment. Cyclotron emission from a two-dimensional electron 
gas heated with an electrical current serves as a magnetic field 
tunable source. The spectrometer is demonstrated at 4.2~K by measuring 
the resonant cyclotron absorption of a second two dimensional electron gas. 
Unique aspects of the spectrometer are that 1) an ultra-broadband
detector is used and 2) the emitter is run
quasi-continuously with a chopping frequency of only 1~Hz. 
Since optical coupling to 
room temperature components is not necessary, this technique is 
compatible with ultra-low temperature (sub 100 mK) operation.

\end{abstract}

}

\section{Introduction}

Many quantum systems of contemporary interest in condensed matter
physics have energy levels in the meV range. These include, but are not
limited to, two dimensional electron gas (2DEG)
systems in high magnetic fields exhibiting the fractional and integer
quantum hall effects, quantum nano-structures such
as quantum dots and carbon nanotubes, and metallic single-electron
transistors. In these systems, temperature dependent behavior
in dc transport measurements is observed all the way down to tens of mK.
The measurement of the far-infrared (THz) transmission through such
systems at ultra-low temperatures could provide complementary
information about the excitation spectrum of the
system.\footnotetext{$^{a)}$Current address: Integrated Nanosystems Research Facility, University of California, Irvine 92697}

The standard technique to measure THz transmission is with a
Fourier Transform Infrared Spectrometer (FTIR) or a molecular gas laser
which couples optically to a cold sample from room temperature.  
The disadvantage of this technique is that broadband, thermal 
blackbody radiation is also coupled to the sample;
this is incompatible with ultra-low 
temperature (sub 100 mK) operation.  Numerically, the (integrated) power 
density is 40~mW/cm$^2$ at 300~K;
typically\cite{MittalWheelerKeller1996} this must be attenuated to 
of order pW in order to avoid sample heating in many ultra-low temperature
systems; at the same time
the radiation in the frequency band of interest must be 
detectable with reasonable signal to noise for the FTIR system to 
operate at all. In principle, this technique can be carried
out. One must construct a set of filters with the desired pass functions from microwave to optical frequencies\cite{BockLange1995}.

In this paper an all-cryogenic spectrometer is presented 
that bypasses the necessity to construct such filters.
The spectrometer is based on  magnetic field tunable
cyclotron resonance (C.R.) emission from a high-mobility
2DEG in a GaAs/AlGaAs heterojunction,
together with a broadband detector. Figure 1 illustrates the concept.
Our demonstration experiment is carried out in a 4~K
environment, but the technique is compatible with
ultra-low temperature (dilution refrigerator) environments. 

A transmission spectrometer using C.R. emission from bulk GaAs was proposed in
1980\cite{Gornik1980}.  Based on that proposal,
a spectrometer was built by Knap\cite{KnapDurRaymond1992} in 1992 using C.R. 
emission from 2DEGs in GaAs, with a spectral resolution of
1.3~cm$^{-1}$ using narrow and moderate band detectors. 
The spectrometer presented here is an improvement 
on that developed by
Knap it two important ways. First, an ultra-broadband detector is used
in place of a narrow band detector. This allows a much wider range of
frequencies to be accessed in principle; in this particular case the
range of 2-65~cm$^{-1}$ are immediately accessible with a single sweep
of the magnetic field. 
Second, we use a very low chopping frequency (1~Hz). Previous
studies of C.R. 
emission\cite{KnapDurRaymond1992,Gornik1981,Gornik1984,Seidenbusch1987,Gornik1984b,vonKlitzing1984,DiesselSiggVonklitzing1991,ChaubetRaymondKnap1991,ZawadzkiChaubetDur1994,ZinovevFletcherChallis1994} typically used microsecond or 
millisecond pulses with low duty cycle for fear of
overheating the sample and inadvertently broadening the emission
linewidth. The experimental results presented here indicate that low
chopping frequencies do not significantly degrade the emission
linewidth. Low chopping frequencies will allow the use of state-of-the-art
ultra-low-noise bolometers, which have
exquisite sensitivity but time constants limited to typically 
tens~of~ms\cite{Bock1998}.  In this paper, we also provide a
quantitative estimate of the generated C.R. power and spectrometer
noise performance, as well as quantitative estimates for the 
ultimate limits of this technique using current as well as future
detector technology with the promise of single photon
sensitivity in the THz band\cite{KomiyamaAstafievAntonov2000,SchoelkopfMoseleyStahle1999,KarasikMcgrathGershenson2000}.

\begin{figure}
\epsfig{file=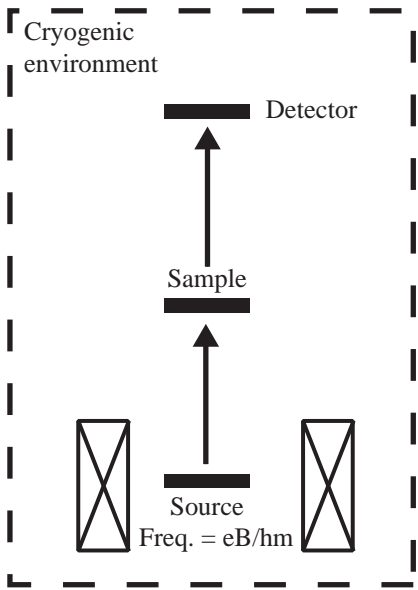}
\caption{Schematic of technique described in this paper.}
\end{figure}

\section{THz source}

\subsection{Principle of operation}

In two-dimensional electron gas in the presence of a magnetic field perpendicular 
to the plane, the spectrum of states (in the absence of disorder 
and interactions) is given by a discrete set of Landau levels, 
with energies given by

\begin{equation}
E=\bigl( n + {1\over{2}} \bigr)~\hbar\omega_c,\\
\end{equation}
where
\begin{equation}
\hbar\omega_c=eB/m^*.\\
\end{equation}
$\omega_c$ is referred to as the cyclotron frequency.
At low temperatures, the lowest energy states are filled, and higher 
energy states are empty. If the electrons are heated to a higher 
temperature, then higher 
energy states are populated. The higher-energy electrons can decay 
to the lower energy levels via phonon emission or photon emission; 
in the latter case the photon frequency is given by $hf~=~eB/m^*$.
In our technique, we heat the electrons with a dc current; the emission 
frequency is then tunable by magnetic field. This cyclotron emission 
from 2DEGs has been studied by many
groups\cite{KnapDurRaymond1992,Gornik1981,Gornik1984,Seidenbusch1987,Gornik1984b,vonKlitzing1984,DiesselSiggVonklitzing1991,ChaubetRaymondKnap1991,ZawadzkiChaubetDur1994,ZinovevFletcherChallis1994,Komiyama1998,KawanoKomiyama2000,KawanoHisanagaKomiyama1999}.
Here, we are mainly interested in characterizing the emission power 
and spectrum in order to use it as a THz source.

Our source consists of a 5x5 mm$^{2}$ (cleaved) 2DEG formed in a 
GaAs/AlGaAs modulation-doped quantum well grown by molecular
beam epitaxy. The mobility and density are $1.25~10^{11}~cm^{-2}$ and  
$600,000~cm^{2}/Vs$, respectively. Ohmic contacts of diffused Au/Ni/Ge 
are deposited along both edges of the sample to make low-resistance 
contacts. We have characterized the emission power using a broadband 
detector. To drive our source electrically, 
we apply a 0.5 Hz sinusoidal voltage to the terminals, typically 
with a 10~k$\Omega$ resistor in series with the emitter. Since heating by 
the electrical current causes spontaneous emission, and since the
power delivered to the emitter from the battery is proportional to
$V^2$, the emission 
power is periodic with a frequency given by twice the electrical drive 
frequency. We thus detect the synchronous voltage on the detector at 
the second harmonic of the drive frequency, 1 Hz. 

The detector is a commercially available broadband composite 
bolometer\cite{HallerBeeman} placed at one end of a 7~''
long 1/2~'' diameter evacuated (gold plated) light pipe. It consists
Ge thermistor mounted on a thin-film (NiCr) coated sapphire absorber
with dimensions 4x4~mm$^2$.
The emitter is placed at the other
end of the light pipe, inside the bore of a superconducting magnet.
The measured 
emission power at the detector for an electrical input power of 1 mW
(electric field of roughly 20~V/cm)
is shown in figure 2 as a function of the source magnetic field. 
The smooth variation of power as a function of magnetic field 
is due to the variation of the 
emitter resistance (and hence the power delivered from the battery)
with the magnetic field.
The periodic structures are interference fringes due to the 
Fabry-Perot etalon formed by the substrate. This effect was modeled
and measured by Zinovev\cite{ZinovevFletcherChallis1994}, 
and provides evidence that the source is
indeed quasi-monochromatic. Three important figures of merit for 
our source are the optical beam pattern, emitted power, and linewidth. 
We discuss these each in turn below.

\subsection{Optics}

The emission from our sample is mostly into the substrate due to its 
high dielectric constant. We have mounted the sample with a thin layer 
of vacuum grease onto a highly-reflective (gold plated) mount. This 
reflects most of the light emitted into the substrate.
We also tried evaporating a thin
metallization layer on the back surface (Al), with no change in the 
output power.  
It is well-known that the emissivity from thin films is distributed 
very uniformly over the entire 2 $\pi$ steridian available solid 
angle\cite{Nishioka1978}. In our case, since the beam 
pattern is spread out even more after reflection and passing through 
the dielectric/air interface, the emission is essentially
isotropically distributed as a function of solid angle.
We use a light-pipe to guide the radiation to the detector; in principle a 
lens system can be used to focus it if that is desired. The radiation 
is circularly polarized for vertical emission, but multiple
reflections off of the light pipe walls and the substrate/sample 
interface at varying angles served to randomize the polarization.

\subsection{Power}

The absolute power available is important, since this sets the 
sensitivity requirements for the detector in our arrangement. 
Absolute power calibrations at this frequency
are typically no better than a factor of
two\cite{DatlaGrossmanHobish1995}, 
and our results for
the measured signal are also reproducible to only a factor of two
between cooldowns. This may be due to drift in the detector
responsivity or to inherent changes in the sample between thermal cycles. 
We estimate the absolute power response of the detector
using the method of dc substitution. In this method, 
we determine the detector responsivity (in V/W) to dc power, and assume the 
responsitivity is the same for THz power. This is 
generally the best agreed upon method for absolute power 
measurements at THz frequencies.  
This gives a responsivity of $3~10^5~V/W$ for our detector; we  
find it also has a noise equivalent power (NEP) of $0.5~pW/\sqrt{Hz}$ at a 
chopping frequency of 1 Hz\cite{NEPnote}. We estimate an optical efficiency 
from the source to the detector of order 10\%; this is mainly 
due to the fact the we use a $1/2"$ diameter light pipe, but the detector 
is only 4x4~mm$^{2}$ in size. (No cone is used to concentrate the radiation
onto the detector.) Thus, from the measured power we determine 
the absolute output power is roughly 100 pW for an electrical 
input power of 1 mW. In a dilution refrigerator environment, the 
emitter would have to be carefully heat sunk to a thermal stage
that could handle the power load. This is possible without too
much difficulty.

Our results for the emission power are consistent with those of 
Zinovev\cite{ZinovevFletcherChallis1994},
who used microsecond electrical pulses and found 
roughly 50 pW of ac power out for roughly 1 mW in, but inconsistent 
with the results of Kawano\cite{KawanoKomiyama2000}, 
who find 50 pW out for 50 mW 
in. These variations in the output power within the literature may 
be related to the sample geometry or the optical coupling 
techniques used.

\subsection{Linewidth}

The spectral resolution of our spectrometer is limited by the
linewidth of the source.
In reality, the spectrum of states in a 2DEG in a magnetic field
is broadened by disorder.
We infer an upper limit on the linewidth of our source by measuring the
transmission vs. magnetic field through a sample with a known resonant
absorption behavior. This is discussed in further detail in 
section III.  Based on these measurements, we infer an upper limit 
of 1.8~cm$^{-1}$ (FWHM) on our source linewidth. These results are consistent
with those of Knap\cite{KnapDurRaymond1992} and
Komiyama\cite{Komiyama1998}, who studied C.R. emission from samples of
similar mobility. The fact that we find such a narrow linewidth even
when running the emitter quasi-continuous is important; this issue is
discussed in more detail below.

If the current distribution and optical emissivity as a function of position
on the surface of the sample were known, it should in principle be possible to 
model the interference fringes
in figure 3 to determine the spectral linewidth. 
Zinovev\cite{ZinovevFletcherChallis1994} modeled the interference fringes for the case of vertical
emission only and found qualitative agreement with the measured fringe pattern.
However, the emission angle is not purely vertical; in reality it is 
spread out over 2~$\pi$ steradians almost evenly (see above).
Since we do not know the details of where
in the sample the emission comes from, it is difficult to relate the
contrast of the interference fringes to the linewidth of the radiation.
Thus, while the presence of fringes 
indicates a component of the emission is monochromatic, it does 
not uniquely determine its linewidth, nor the amount of incoherent 
broadband radiation. The fringe pattern did not vary strongly with
the mobility of the emitter sample 
($\mu~=~600,000~cm^{2}/Vs$ vs. $2.7~10^6~cm^2/Vs$), suggesting that the 
``contrast'' of the interference fringes
is not a good measure of the emission linewidth.

\begin{figure}
\epsfig{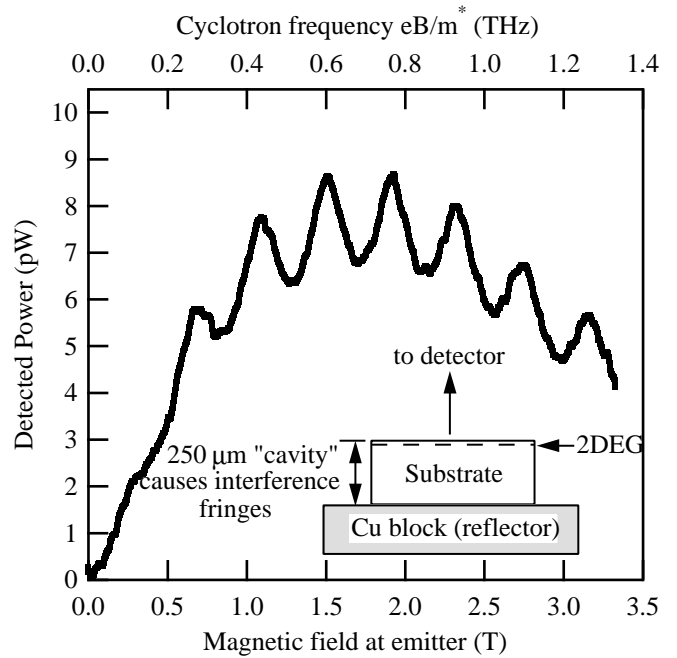}
\caption{Detected power vs. magnetic field.}
\end{figure}

\subsection{Chopping technique}

In contrast to most C.R. emission experiments,
we run our sources quasi-continuous, and not pulsed, and still find reasonably 
narrow linewidth compared to the results found in the literature. 
This an important point if one wants to use this technique with
ultimate state-of-the-art low-noise 
detectors. Fast detectors used for pulsed measurement typically 
have NEPs of
$10^{-12}~W/\sqrt{Hz}$\cite{BrownWenglerPhillips1985,StrasserBochterWitzany1991}.
With the output power of our source 
in the 10-100~pW range, this gives limited signal to noise. Recent 
progress on ultra-low noise detectors for measuring the cosmic 
microwave background have achieved NEPs of 
2~10$^{-18}~W/\sqrt{Hz}$\cite{Bock1998} at an operating temperature of 100~mK.
Although ultra-sensitive, they 
have time constants (typically 10 ms) that do not allow for pulsed 
operation. Thus, our quasi-continuous technique is important if those ultra-low 
noise detectors are to be used.

\begin{figure}
\epsfig{file=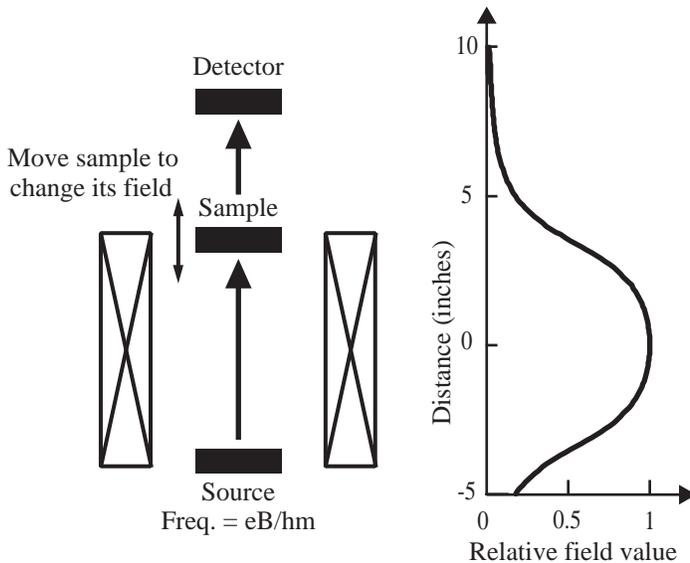}
\caption{Actual setup.}
\end{figure}

Initial work on  {\it fast} low noise detectors based on
lithographically fabricated hot-electron ``microbolometers'' began in the late
1980s\cite{GershenzonGershenzonGoltsman1989}.
In a later variant of the idea\cite{Nahum1993},
electrical NEPs of $10^{-17}~W/\sqrt{Hz}$ with a response time
constant of 10~$\mu$s were measured. If these could be 
antenna coupled, they would make
fast, sensitive detectors. 
This would obviate the need for the low
chopping frequencies described in this paper. 
This general line of investigation is 
currently being pursued by several groups 
in order to demonstrate optical NEPs in that 
range or better with fast response
time\cite{SchoelkopfMoseleyStahle1999,KarasikMcgrathGershenson2000,CatesBricenoSherwin1998,Hergenrother1995}.
Nonetheless, our results
show that the development of fast, sensitive detectors is not a
necessity for the technique described herein. 
Current state-of-the-are ultra-sensitive bolometer technology can be used 
without a degradation in the performance (i.e. linewidth) of the
spectrometer due to the low chopping frequency.

\section{THz Transmission Spectrometer}

\subsection{Principle of operation}

In principle, our technique is simple and straightforward. The
frequency of the source is tuned by its local magnetic field, and 
the light is guided through the sample and onto the detector. One 
can sweep the source frequency by sweeping its magnetic field; the 
power measured on the detector is a direct measure of the transmission 
through the sample at each frequency. An alternative mode 
of operation is to fix the source frequency (by fixing it's local 
magnetic field) and sweep some sample parameter, to determine the 
transmission at a fixed frequency vs. the sample parameter.
We discuss this in more detail in the next section.

\subsection{Demonstration of operation}

We have carried out a demonstration experiment to measure the
transmission through another 2DEG at a fixed frequency as a function 
of magnetic field at the sample. For a 2DEG in a magnetic field,
absorption of THz photons occurs only if photon energy matches
the cyclotron energy, $hf=eB/m^*$. Thus, as a function of frequency of
the emitter (or magnetic field of the sample as realized here),
the transmission is unity (neglecting the vacuum-dielectric mismatch)
off resonance, and minimum on resonance. 
If the light is randomly polarized, the absorption coefficient depends
on the sample mobility and density. The transmission minimum
is limited theoretically to 50\%, although weaker absorption 
is usually observed.

In our spectrometer, the light from the emitter passes through 
a $1/2$" diameter light pipe for about 3", and is then 
passed through a 3 degree cone onto the transmission sample 
through a 1/8" hole, and then through another $1/2$'' light
pipe onto the detector. Instead of using two separate
magnets, one to set the emitter frequency and one to set the sample
magnetic field, we use a one-magnet setup indicated schematically in 
figure 3. By fixing the current through the 
superconducting magnet, we fix the magnetic field at the source
and hence the emission frequency. (The emitter is 
located towards the bottom of the magnet at the 55~\% field region.)
We then monitor the detector voltage while sweeping the magnetic 
field at the sample; this is accomplished by physically moving 
the sample in the region where the field strength varies with position.
The field profile is shown in figure 3; by moving the sample
up and down two inches, we can vary the sample magnetic field from
25\% to 75\% of the value at the field center. 
With a 9~T superconducting magnet,
we can achieve a maximum field of 5~T at the emitter, corresponding to
a maximum emission frequency of 65~cm$^{-1}$.

We plot in figure 4 the measured transmission coefficient as a
function of sample magnetic field.
The right axis is the measured power on the detector;
the emitter frequency is fixed at 1.1 THz. 
We find a Lorentzian resonance with a FWHM linewidth of
0.26~Tesla (2.7~cm$^{-1}$) for this sample 
($\mu=600,000~cm^2/V-s$).
On another sample with somewhat higher mobility ($2.7~10^6~cm^2/V-s$),
we measure a linewidth of roughly 0.2~Tesla (1.8~cm$^{-1}$). 
We conclude from this that the emitter 
linewidth is no broader than 1.8~cm$^{-1}$; 
otherwise the sharp absorption feature 
in figure 4 would not be seen. 

It is possible that there is a spectral component of the emitted power
which is broadly distributed away from $\omega_C$.
From the measurement described, we can set an upper limit 
on the amount of this component. Since the absorption for
randomly polarized light is at most 50\% and our absorption dip is
about 10\%, we conclude that at most 80\% of the emitted 
power of the source is off resonance; and that at least 20\% power is 
concentrated in the spectral region around $\omega_c$. However, this
is only an upper limit on the out of band emission
because we have not independently measured the transmission of our
sample; if the true sample absorption is less than the theoretical limit of
50\% on resonance, then the background component of the emission is
less than 80\%.

\begin{figure}
\epsfig{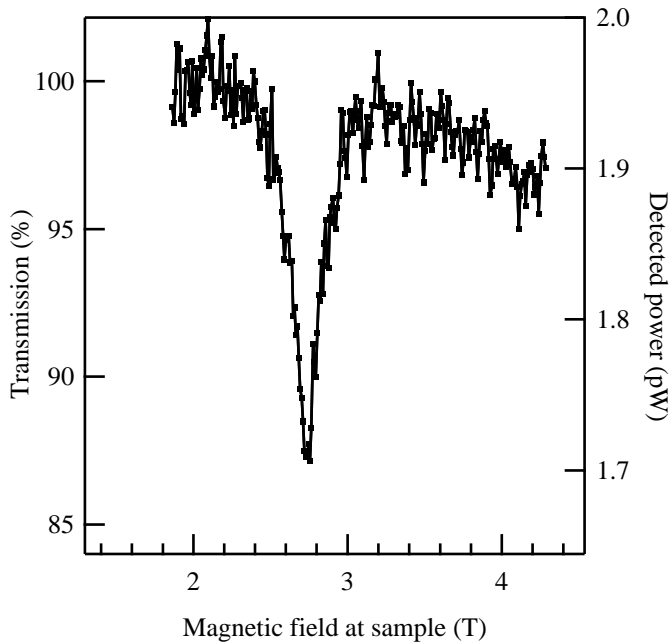}
\caption{Resonant absorption. The emitter field is fixed at 2.75~T;
  hence the emission frequency is fixed at 1.1~THz.}
\end{figure}

The linewidth of cyclotron resonance absorption in high mobility 2DEGs
has been studied for many years by many groups. A very typical
feature is that the linewidth varies periodically with filling factor
$\nu=nh/eB$, and that the linewidth is an absolute maximum at filling
factor 2. Our measurements presented here are near $\nu~=~2$, so
that our measurement of the spectral resolution is probably an upper
limit on the ultimate attainable spectral resolution using this technique.

In figure 6, we plot the measured linewidth vs. sample mobility 
at a filling factor of $\nu=2$, as well as several
measured values from the
literature\cite{SchlesingerAllenHwang1984,SeidenbuschGornikWeimann1987,BatkeStormerGossard1988,NicholasHopkinsBarnes1989,EnglertMaanUihlein1983,HeronLewisClark2000}. The temperature for the references varied
between 1.3~K and 5~K. The linewidths we measure
are comparable to those measured by other groups, suggesting
that our emitter linewidth is narrow enough to measure roughly the
correct value for the transmission linewidth.

We have also measured the 
transmission through the sample off of the resonant absorption as 
a function of source frequency. We find roughly unity transmission 
off resonance. We have also carried out this swept frequency
transmission measurement though a blank sample (containing no 2DEG) of
GaAs. In either case, we do not observe any Fabry-Perot fringes due to 
the finite (transmitting) sample thickness, even though 
the sample is not wedged. 
We conclude from this that the incoming radiation is equally 
distributed as a function of angle of incidence. This is an added 
advantage of this technique; the transmission 
samples do not need to be wedged. 

Finally, even though the emitter must be placed in a magnetic field,
many dilution refrigerators have compensating coils to minimize the
magnetic field at the mixing chamber. By placing the transmission
sample and detector near this compensated region, the technique
described here can be applied to systems in low magnetic fields
as well.

\subsection{Ultimate sensitivity limits}

Based on the noise of our detector, we can measure the power
transmitted through the sample with a statistical uncertainty of
roughly 0.5~pW in a one second integration time. In figure 4, the
transmitted power is roughly 2~pW off resonance; we used an equivalent
noise bandwidth of approximately 1~mHz for that measurement. The statistical
error on each transmission measurement point is thus about 1~\%.
This noise performance is clearly marginal, and underscores 
the need for lower noise detector
technology. If the spectrometer described in this paper were to be
incorporated into a low temperature environment, low-noise state of
the art bolometers\cite{Bock1998} 
with NEPs of as low as 2~10$^{-18}~W/\sqrt{Hz}$
could be used. This would allow for a much better measurement of the
transmission coefficient, with a statistical uncertainty of
10$^{-5}$~\%, given the power levels we use.

For a given statistical uncertainty in the measured transmission
coefficient at each frequency, we can predict the statistical
uncertainty on the measured linewidth and line position for a
Lorentzian absorption profile. Based on numerical simulations of
typical experimental parameters, we find a statistical uncertainty in
the measured linewidth of roughly 25 Gauss (or 1~GHz in frequency) per
1\% error in measured transmission coefficient, and roughly 2.5~Gauss
(0.1~GHz) in the measured position per 1\% error in measured transmission
coefficient. With a detector with NEP of 2~10$^{-18}~W/\sqrt{Hz}$,
we could determine the linewidth and center
frequency with a statistical uncertainty of order 1~kHz.
We note that if this technique were used on samples with
narrower absorption features than the emitter linewidth, 
it would not be possible
to determine the {\it shape} of those absorption feature.
However, it would be possible to
determine the {\it position} (in frequency or magnetic field) 
of the absorption feature to a precision much better than the emitter
linewidth.

\begin{figure}
\epsfig{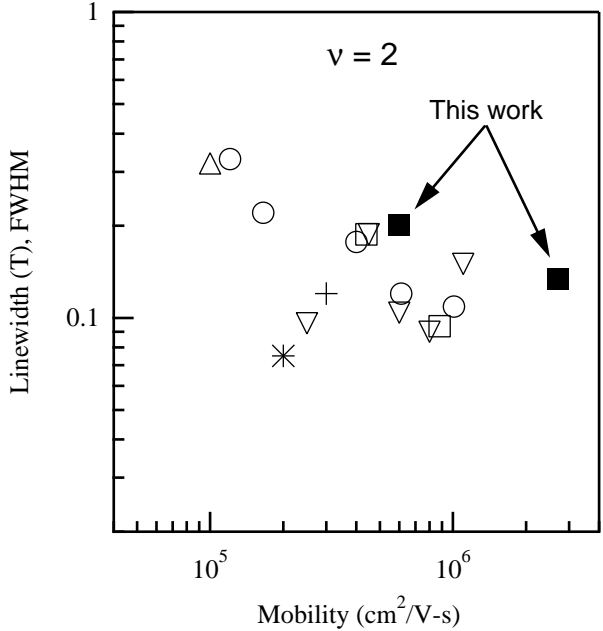}
\caption{Linewidth vs. mobility for this and other references, at
  $\nu$=2. Symbols are: $\ast$ Schlesinger 1984, 
$\bigcirc$ Seidenbusch 1987, square Batke 1988,$\bigtriangledown$ 
Nicholas 1989, $\bigtriangleup$ Englert 1983, $+$ Heron 2000.} 
\end{figure}

Recent work on THz detectors has demonstrated single-photon
sensitivity\cite{KomiyamaAstafievAntonov2000} with unknown
quantum-efficiency and proposals
exist\cite{SchoelkopfMoseleyStahle1999,KarasikMcgrathGershenson2000} 
for single-photon sensitivity with excellent predicted quantum
efficiency.  These detectors must be operated in an ultra-low
temperature environment which is very compatible with the spectrometer
presented here. We now consider the ultimate limits of the
spectrometer developed here if single photon THz detectors were to be used.

At low (dilution refrigerator) temperatures, the thermal background
of THz photons is negligible. 
Therefore, the statistical uncertainty in the measured
signal would be limited only by the ``shot'' noise of the incoming
photons. Roughly $10^9$ photons/second are
generated in a 1~pW beam. For {\it classical} statistics, the fluctuations 
are $\sqrt{10^9}$ photons/second, i.e. $10^{-17}$ W; for {\it quantum}
statistics the fluctuations can be lower or higher, corresponding to
bunching or antibunching\cite{MandelandWolf:1995}. Generally, the
statistics of the radiation depends on the emission process.
To our knowledge this has not been investigated (either theoretically or 
experimentally) in cyclotron resonance emission. These calculations
suggest that perfect (noiseless) THz detectors would not significantly
improve the noise performance of our spectrometer, as compared to
state-of-the-art low noise detectors with NEPs of
2~10$^{-18}~W/\sqrt{Hz}$, since the statistics of the signal dominate
the noise performance. However, the ultimate limit of 
a few photons emitted per
second with a single photon THz detector, where the photons interact
with a quantum system of interest between the source and detector,
may give experimentalists new
tools to explore quantum information processing in condensed matter
systems\cite{SherwinImamogluMontroy1999,ColeWilliamsKing2001}.

\section{Conclusions}

Using a 2DEG as a cyclotron resonance source, 
we have developed an all-cryogenic THz transmission spectrometer for the
spectral range of 2-65~cm$^{-1}$, with a spectral resolution of at
least 1.8~cm$^{-1}$. Since a broadband detector is used, 
the entire frequency range can be accessed by the single sweep of the
magnetic field at the emitter. Additionally, we have demonstrated that a low
chopping frequency does not degrade the emission linewidth; this is
important for future use with ultra-low noise detectors. The
spectrometer was demonstrated by measuring the 
THz transmission through another 2DEG as a function of magnetic field of 
the 2DEG. 

\section{Acknowledgements}

This work was supported by Sandia National Labs Grant No. DE-AC04-AL85000
 and DOE Grant No. DE-FG03-99ER45766. One of the authors (P.J.B.) 
was supported in part by the Sherman Fairchild Foundation.


\begin{references}
\bibitem{NEPnote}
The bolometer we used had normal metal (0.0005" diameter) brass leads for the
  electrical readout. If superconducting electrical leads would have been used,
  a lower NEP could have been achieved. We chose normal metal leads for this
  prototype experiment because we did not use a compensating coil at the
  detector to null the magnetic field; stray magnetic fields at the detector
  could have affected its responsivity if we used superconducting leads.
\end{references}

\end{document}